\documentclass[%
 reprint,
superscriptaddress,
 amsmath,amssymb,
 aps,
prl,
]{revtex4-1}

\usepackage{graphicx}% Include figure files
\usepackage{dcolumn}% Align table columns on decimal point
\usepackage{bm}% bold math
\usepackage{color}

\begin{document}

%\preprint{APS/123-QED}

\title{Photo-induced plasmon-phonon coupling in PbTe}% Force line breaks with \\

\author{M.P. Jiang}
 \affiliation{Stanford PULSE Institute, SLAC National Accelerator Laboratory, Menlo Park, CA 94025, USA}
 \affiliation{Stanford Institute for Materials and Energy Sciences, SLAC National Accelerator Laboratory, Menlo Park, CA 94025, USA}
 \affiliation{Department of Physics, Stanford University, Stanford, CA 94305, USA}
\author{M. Trigo}
 \affiliation{Stanford PULSE Institute, SLAC National Accelerator Laboratory, Menlo Park, CA 94025, USA}
 \affiliation{Stanford Institute for Materials and Energy Sciences, SLAC National Accelerator Laboratory, Menlo Park, CA 94025, USA}
\author{S. Fahy}
 \affiliation{Tyndall National Institute and Department of Physics, University College, Cork, Ireland}
\author{A. Hauber}
 \affiliation{Tyndall National Institute and Department of Physics, University College, Cork, Ireland}
\author{\'{E}.D. Murray}
 \affiliation{Tyndall National Institute and Department of Physics, University College, Cork, Ireland}
 \affiliation{Departments of Physics and Materials, Imperial College London, London SW7 2AZ, UK}
\author{I. Savi\'{c}}
 \affiliation{Tyndall National Institute and Department of Physics, University College, Cork, Ireland}
\author{C. Bray}
 \affiliation{Stanford PULSE Institute, SLAC National Accelerator Laboratory, Menlo Park, CA 94025, USA}
 \affiliation{Department of Applied Physics, Stanford University, Stanford, CA 94305, USA}
\author{J. N. Clark}
 \affiliation{Stanford PULSE Institute, SLAC National Accelerator Laboratory, Menlo Park, CA 94025, USA}
\author{T. Henighan}
 \affiliation{Stanford PULSE Institute, SLAC National Accelerator Laboratory, Menlo Park, CA 94025, USA}
 \affiliation{Stanford Institute for Materials and Energy Sciences, SLAC National Accelerator Laboratory, Menlo Park, CA 94025, USA}
 \affiliation{Department of Physics, Stanford University, Stanford, CA 94305, USA}
\author{M. Kozina}
 \affiliation{Stanford PULSE Institute, SLAC National Accelerator Laboratory, Menlo Park, CA 94025, USA}
 \affiliation{Stanford Institute for Materials and Energy Sciences, SLAC National Accelerator Laboratory, Menlo Park, CA 94025, USA}
 \affiliation{Department of Applied Physics, Stanford University, Stanford, CA 94305, USA}
\author{M. Chollet}
 \affiliation{Linac Coherent Light Source, SLAC National Accelerator Laboratory, Menlo Park, CA 94025, USA}
\author{J.M. Glownia}
 \affiliation{Linac Coherent Light Source, SLAC National Accelerator Laboratory, Menlo Park, CA 94025, USA}
\author{M.C. Hoffmann}
 \affiliation{Linac Coherent Light Source, SLAC National Accelerator Laboratory, Menlo Park, CA 94025, USA}
\author{D. Zhu}
 \affiliation{Linac Coherent Light Source, SLAC National Accelerator Laboratory, Menlo Park, CA 94025, USA}
\author{O. Delaire}
 \affiliation{Department of Mechanical Engineering and Materials Science, Duke University, Durham, NC 27708, USA}
 %\affiliation{Materials Science and Technology Division, Oak Ridge National Laboratory, Oak Ridge, Tennessee 37831, USA}
\author{A.F. May}
 \affiliation{Materials Science and Technology Division, Oak Ridge National Laboratory, Oak Ridge, Tennessee 37831, USA}
\author{B.C. Sales}
 \affiliation{Materials Science and Technology Division, Oak Ridge National Laboratory, Oak Ridge, Tennessee 37831, USA}
\author{A.M. Lindenberg}
 \affiliation{Stanford PULSE Institute, SLAC National Accelerator Laboratory, Menlo Park, CA 94025, USA}
 \affiliation{Stanford Institute for Materials and Energy Sciences, SLAC National Accelerator Laboratory, Menlo Park, CA 94025, USA}
 \affiliation{Department of Materials Science and Engineering, Stanford University, Stanford, CA 94305, USA}
\author{P. Zalden}
 \affiliation{Stanford PULSE Institute, SLAC National Accelerator Laboratory, Menlo Park, CA 94025, USA}
 \affiliation{Stanford Institute for Materials and Energy Sciences, SLAC National Accelerator Laboratory, Menlo Park, CA 94025, USA}
 \affiliation{Department of Materials Science and Engineering, Stanford University, Stanford, CA 94305, USA}
\author{T. Sato}
 \affiliation{Linac Coherent Light Source, SLAC National Accelerator Laboratory, Menlo Park, CA 94025, USA}
 %\affiliation{RIKEN SPring-8 Center, Kouto 1-1-1, Sayo, Hyogo 679-5148, Japan}
 %\affiliation{Department of Chemistry, The School of Science, The University of Tokyo, 7-3-1 Hongo, Bunkyo-ku, Tokyo 113-0033, Japan}
\author{R. Merlin}
 \affiliation{Department of Physics, University of Michigan, Ann Arbor, Michigan 48109, USA}
\author{D.A. Reis}
 \affiliation{Stanford PULSE Institute, SLAC National Accelerator Laboratory, Menlo Park, CA 94025, USA}
 \affiliation{Stanford Institute for Materials and Energy Sciences, SLAC National Accelerator Laboratory, Menlo Park, CA 94025, USA}
 \affiliation{Department of Applied Physics, Stanford University, Stanford, CA 94305, USA} 

\date{\today}% It is always \today, today,

\begin{abstract}
We report the observation of photo-induced plasmon-phonon coupled modes in the group IV-VI semiconductor PbTe using Fourier-transform inelastic X-ray scattering at the Linac Coherent Light Source (LCLS). We measure the near-zone-center dispersion of the heavily screened longitudinal optical (LO) phonon branch as extracted from differential changes in x-ray diffuse scattering intensity following above band gap photoexcitation. 
\end{abstract}

\maketitle

In polar semiconductors Fr\"{o}hlich electron-phonon interactions lead to strong coupling between collective electronic and longitudinal lattice excitations. The coupling can be pronounced in the group IV-VI compounds due to the combination of high polarizability and large longitudinal optical (LO) / soft transverse optical (TO) phonon splitting near zone center. This leads to rapidly dispersing LO phonon--plasmon coupled (LOPC) modes \cite{cowley1965prl,pawley1966prl,varga1965pr,kim1978prb} that can affect nonequilibrium properties such as carrier relaxation\cite{PhysRevB.37.6290} and transport\cite{KOBAYASHI1975875}, and have implications for superconductivity at high carrier densities\cite{allen1969carrier}. Inelastic neutron scattering (INS) measurements\cite{cowley1965prl,alperin1972pla,cochran1966proc,pawley1966prl,elcombe1967proc} on several doped group IV-VI semiconductors, including PbTe, show an anomalous dip in the low wavevector/long-wavelength dispersion of the LO phonon branch, due to screening from from the free carrier concentrations.
{\par}

Ultrafast photoexcitation can be used to transiently control materials properties, for example,  through the excitation of large amplitude vibrational motion 
or the generation of large carrier densities exceeding 10$^{20}$cm$^{-3}$. Such excitation can furthermore lead to a nonequilibrium state with dramatically different properties from the ground state. Photoexcited LO-phonon--plasmon coupled modes have been observed in all-optical experiments in the III-V compound GaAs\cite{ishioka2011prb,hase1999prb,cho1996prl} as well as PbTe \cite{Wu2007} and PbTe$_{0.95}$S$_{0.5}$\cite{Romcevic2004}.
We have previously shown using time-resolved x-ray diffuse scattering that near band-gap photoexcitation in PbTe couples the TO and transverse acoustic (TA) modes at high wavevector along the bonding direction, reducing the ferroelectric instability and stabilizing the paraelectric state\cite{jiang2016natcomm}. In this case, photoexcitation is also expected to strongly affect the LO phonon through coupling to the photoexcited plasma. Here we present femtosecond Fourier-transform inelastic X-ray scattering (FT-IXS)\cite{trigo2013naturephys,zhu2015prb,henighan2016prb,jiang2016natcomm} measurements of near zone center excitations in photoexcited PbTe.
In this technique, absorption of a long-wavelength pump pulse produces correlated pairs of phonons with equal and opposite momenta, leading to squeezing oscillations in the mean-square phonon displacements. The subsequent dynamics are probed  by femtosecond x-ray diffuse scattering.
We observe a heavily damped mode that strongly disperses with increasing wavevector from near the zone center TO frequency to the LO frequency. We attribute the time- and wavevector-dependent  signal to squeezed oscillations of the LOPC mode, likely coupled to low energy single-particle excitations.{\par} 

{\par}

	The experiment was performed at the X-ray pump probe (XPP) instrument \cite{chollet2015jsync} of the Linac Coherent Light Source (LCLS) X-ray free-electron laser. Details of the experimental setup can be found in reference \cite{jiang2016natcomm}. Briefly, infrared pulses of light (60 fs, 350 $\mu$J, 0.6 eV) generated from an optical parametric amplifier laser were used as the pump source and hard x-ray pulses (50 fs, 8.7 keV) as the probing mechanism. The energy of the pump source was chosen to just exceed the direct band gap of PbTe ($\sim$0.31 eV at room temperature\cite{sitter1977prb}). A large area Cornell-SLAC Pixel Array Detector (CSPAD) captured the resulting x-ray diffuse scattering over a wide region of reciprocal space. Diffuse scattering patterns were recorded at room temperature tracked as a function of time delay $\tau$ between the IR pump and x-ray probe pulses at binned step sizes of 100 fs.{\par}
	We chose a fixed sample and detector configuration such that we capture scattering with a momentum transfer near the $(\bar{1} 1 3)$ Brillouin zone. The sample was detuned $\sim$1\% (in rlu) from the Bragg condition for reciprocal lattice vector $\mathbf{G}=(\bar{1} 1 3)$ to prevent the full intensity of the Bragg reflection from hitting the detector. Two high-symmetry reduced wavevector directions are captured (approximately): $\Gamma$ towards X with $\mathbf{q}\sim (0 q_{y} 0)$ and $\Gamma$ towards W with $\mathbf{q}\sim (q_{x} 0 q_{z}=2 q_{x})$. Notably, in our measurement scheme, the two directions have varying sensitivity to phonon polarization due to the $(\mathbf{Q}\cdot \mathbf{e})^2$ dependence in the scattering intensity where $\mathbf{e}$ is the phonon polarization, and $\mathbf{Q}=\mathbf{G}+\mathbf{q}$ is the momentum transfer (divided by $\hbar$). Thus, the diffuse scattering along $\Gamma$ to X is primarily sensitive to phonons of transverse polarization (along $z$). Conversely, the diffuse scattering along $\Gamma$ to W is largely sensitive to phonons of longitudinal polarization. Thus, the observation of the screened LO phonon branch is not expected along $\Gamma$ to X. 
	
\begin{figure}
\centering
\includegraphics[width=1\linewidth]{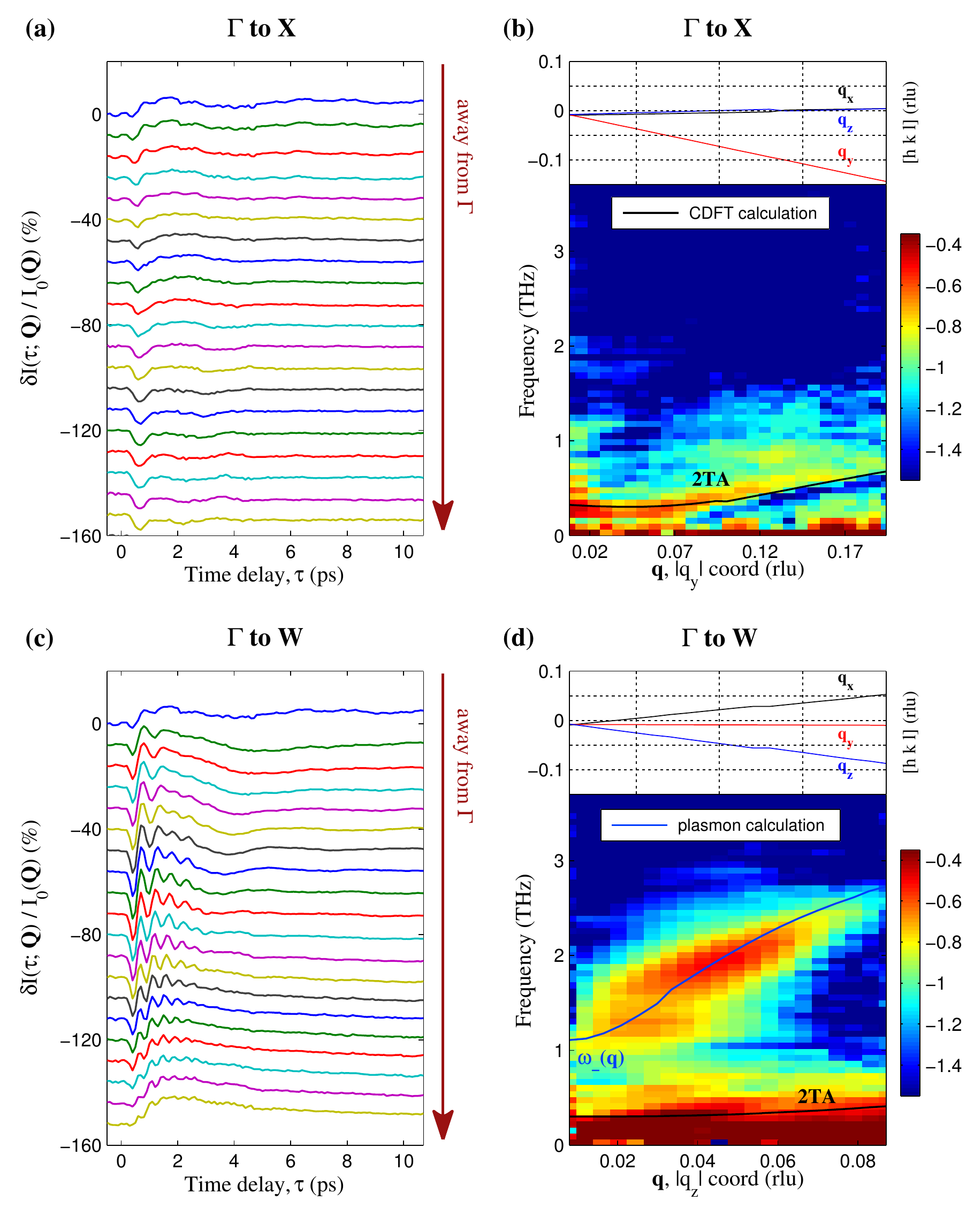}
\caption{\label{f:fig2}(a),(c) Time-domain traces of relative differential diffuse scattering intensities extracted from pixels along the $\Gamma$ to X and $\Gamma$ to W wavevector directions respectively in the $(\bar{1} 1 3)$ BZ. The topmost traces represent wavevector coordinates closest to $\Gamma$ whereas the bottommost traces represent coordinates furthest from $\Gamma$. (b),(d) FT-IXS spectra along the (b) $\Gamma$ to X and (d) $\Gamma$ to W directions as extracted from the amplitudes following Fourier-transform analysis. The amplitudes are displayed on a base-10 logarithmic scale. The coordinate values $(q_{x} q_{y} q_{z})$ for the reduced wavevector directions are plotted above the spectra. Black traces in the spectra are two-phonon dispersion frequencies from first principles calculations. The blue trace in (d) is the calculated approximate trajectory of the screened LO phonon branch $\omega_{-}(\mathbf{q})$ in the long-wavelength limit.}
\end{figure}

	The differential scattering intensity $\delta I(\tau;\mathbf{Q}) = I(\tau;\mathbf{Q}) - I_{0}(\mathbf{Q})$ are collected as a function of time-delay, $\tau$. $I_{0}$ is the unpumped signal, collected for  time delays where the x-ray probe arrives prior to the pump pulse ($\tau< 0$).
    The behavior along the two high-symmetry directions described above are shown directly in the time-domain in Figure \ref{f:fig2}. Figures \ref{f:fig2}(a) and \ref{f:fig2}(c) show extracted $\delta I(\tau;\mathbf{Q})$ values along the respective $\Gamma$ to X and $\Gamma$ to W wavevectors, relative to $I_{0}(\mathbf{Q})$. Each trace  represents the time-dependent differential changes of a pixel on the detector, and thus a unique $\mathbf{Q}$ along one of the labeled high-symmetry wavevectors. The topmost traces depict data nearest to zone center. The traces extracted along the $\Gamma$ to X wavevector reveal a strong decrease in scattering intensity immediately following the arrival of the pump pulse and weak modulations with long periods. The activity along $\Gamma$ to X across the entire BZ has already been described in a prior report \cite{jiang2016natcomm}. Along the $\Gamma$ to W wavevector, on the other hand, substantial coherent oscillations with much faster periods are present, damping on a sub-picosecond timescale. Moreover, the oscillation periods shorten as the traces denote $\mathbf{q}$ coordinates that reside further from $\Gamma$, an indication of dispersive behavior.{\par} 

	A sudden change in the interatomic force constants before and after photoexcitation leads to the time evolution of the displacement correlations between phonon modes. The time-dependent diffuse x-ray signal is proportional to those displacement correlations\cite{trigo2013naturephys,fahy2016prb} and its Fourier transform gives an FT-IXS spectrum. We plot the FT-IXS spectra\cite{trigo2013naturephys,zhu2015prb,henighan2016prb} of the collected time-domain data in Figs. \ref{f:fig2}(b) and \ref{f:fig2}(d) for a range of $\mathbf{q}$ from near $\Gamma$ towards X and W respectively. The amplitudes are displayed in false color on a  logarithmic (base-10) scale  as a function of $|q_{y}|$ in Fig. \ref{f:fig2}(b) and $|q_{z}|$ for \ref{f:fig2}(d). The complete trajectories of $\textbf{q}$ coordinate values for both wavevectors are displayed in rlu above the spectra.
	
	We identify features in the FT-IXS spectra with the help of first principles calculations \cite{murray2005prb,he2012pccp,fahy2016prb,jiang2016natcomm} of the displacement correlations between phonon modes due to a sudden promotion of valence electrons to the conduction band. The full details of the calculations can be found in reference \cite{jiang2016natcomm}. In Fig. \ref{f:fig2}(b), an overtone transverse acoustic mode (2TA) is identified as shown by the black trace juxtaposed over the $\Gamma$ towards X spectrum. The appearance of this mode near $\Gamma$ is consistent with the results discussed in ref. \cite{jiang2016natcomm} in which the same mode is identified along $\Gamma$ to X, all the way out to zone edge.{\par}
    While the spectral feature along the $\Gamma$ to X direction can be identified with the aid of first principles calculations, this model does not identify the vivid feature along the $\Gamma$ towards W spectrum. The calculated excited-state dispersion for the overtone TA mode is overlaid as a black trace in the spectrum of Fig. \ref{f:fig2}(d) and agrees well with the low-frequency feature near $\sim$0.25 THz. Note that although this specific wavevector direction is primarily sensitive to longitudinal phonon polarizations, the overtone TA branch still appears due to a non-negligible residual sensitivity to transverse polarizations. The intense and broad highly-dispersive feature we identify as the LOPC mode as described below.{\par}

    	The interaction between plasmons and the LO phonon mode in polar materials \cite{cowley1965prl,pawley1966prl,varga1965pr,kim1978prb} screens the macroscopic electric field associated with the LO branch, reducing its strength by a factor equaling the low frequency dielectric constant $\varepsilon(\mathbf{q},0)$. In the long-wavelength limit, the coupling is largest when the zone center plasma frequency $\omega_{p}(\mathbf{q}=0)$ equals the LO frequency $\omega_{LO}(\mathbf{q}=0)$. Here, the plasma frequency depends on the  concentration of free carriers $n_{c}$, their charge and effective mass $q_{e}$ and $m^{*}$, the vacuum permittivity $\varepsilon_{0}$, and the high-frequency dielectric constant $\varepsilon_{\infty}$ as 
	\begin{equation*}\omega_{p}^{2}(\textbf{q}=0) = \frac{n_{c}q_{e}^{2}}{m^{*}\varepsilon_{0}\varepsilon_{\infty}}.
	\end{equation*}
 For group IV-VI semiconductors, $\omega_{p}(\textbf{q}=0)$ $\sim$ $\omega_{LO}(\textbf{q}=0)$ for carrier concentrations as small as $\sim$10$^{17}$ cm$^{-3}$\cite{kim1978prb}.{\par}
    At higher carrier densities, when $\omega_{p}(\mathbf{q}=0)$ exceeds $\omega_{LO}(\mathbf{q}=0)$, the dielectric function can be approximated in the quasistatic limit ($\omega \rightarrow$ 0) as $\varepsilon(\textbf{q},0)$.
For $q \ll k_s$, the electric field of the LO mode is dramatically screened leading to dispersion in the lower frequency coupled plasmon--LO phonon  \cite{cowley1965prl}, 
\begin{equation}
\omega^2_{-}(\textbf{q}) = \omega_{TO}^{2}(\textbf{q}=0) + \frac{\omega_{LO}^{2}(\textbf{q}=0)-\omega_{TO}^{2}(\textbf{q}=0)}{\varepsilon(\textbf{q},0)},
\end{equation}
where $\omega_{TO}(\textbf{q}=0)$ is the zone center TO phonon frequency ($\sim$0.95 THz for PbTe at room temperature)\cite{cochran1966proc}.{\par} 
	
	For large enough carrier density, such as in dense photoexcitation,  the Fermi energy well exceeds the thermal energy and we can approximate the carriers as a degenerate Fermi gas. In this case we approximate the fermi-wavevector within a single L-valley as,
	$k_{F} = (3\pi^{2} (n_{c}/4))^{1/3}$.

This leads to a Thomas-Fermi screening wavevector\cite{ashcroftmerminbook},
\begin{equation}
k_{s} = \frac{q_{e}}{\hbar}\left(\frac{3n_{c}m^{*3}}{(\varepsilon_{0}\varepsilon_{\infty})^{3}\pi^{4}}\right)^{{1}/{6}}.
\label{e:ks}\end{equation}.

In our experiments, strong plasmon-phonon coupling is expected from the initially low-carrier-density (4$\times$10$^{17}$ cm$^{-3}$) n-type PbTe upon photoexcitation of $\sim 2 \times 10^{20} \text{ cm}^{-3}$ carriers per L-valley above the band gap ($\sim 0.5$ \% valence excitation). In this excitation regime, the frequency of the photoexcited plasma $\omega_{p}(\mathbf{q}=0)$ ($\sim$73 THz) far exceeds $\omega_{LO}(\mathbf{q}=0)$ ($\sim$3.42 THz for PbTe)\cite{cochran1966proc}. 
Here k$_{F} \sim 0.186 $ reciprocal lattice units (rlu) and k$_{s} \sim 0.083$ rlu respectively. The use of a degenerate Fermi gas at 0 K model to compute these parameters is valid here since the approximate Fermi energy ($\sim$1.3 eV) far exceeds the room temperature thermal energy, considering the estimated photoexcited carrier density. Note, however, that the particular details of this model do not have a significant impact on the forthcoming interpretation of our experimental results. For instance, the stated photoexcited carrier density is strictly an estimate and the exact quantity does not strongly influence the results due to the weak $n_{c}^{1/6}$ dependence for the screening wavevector seen in eq.(\ref{e:ks}).{\par}

\begin{figure}
\centering
\includegraphics[width=1\linewidth]{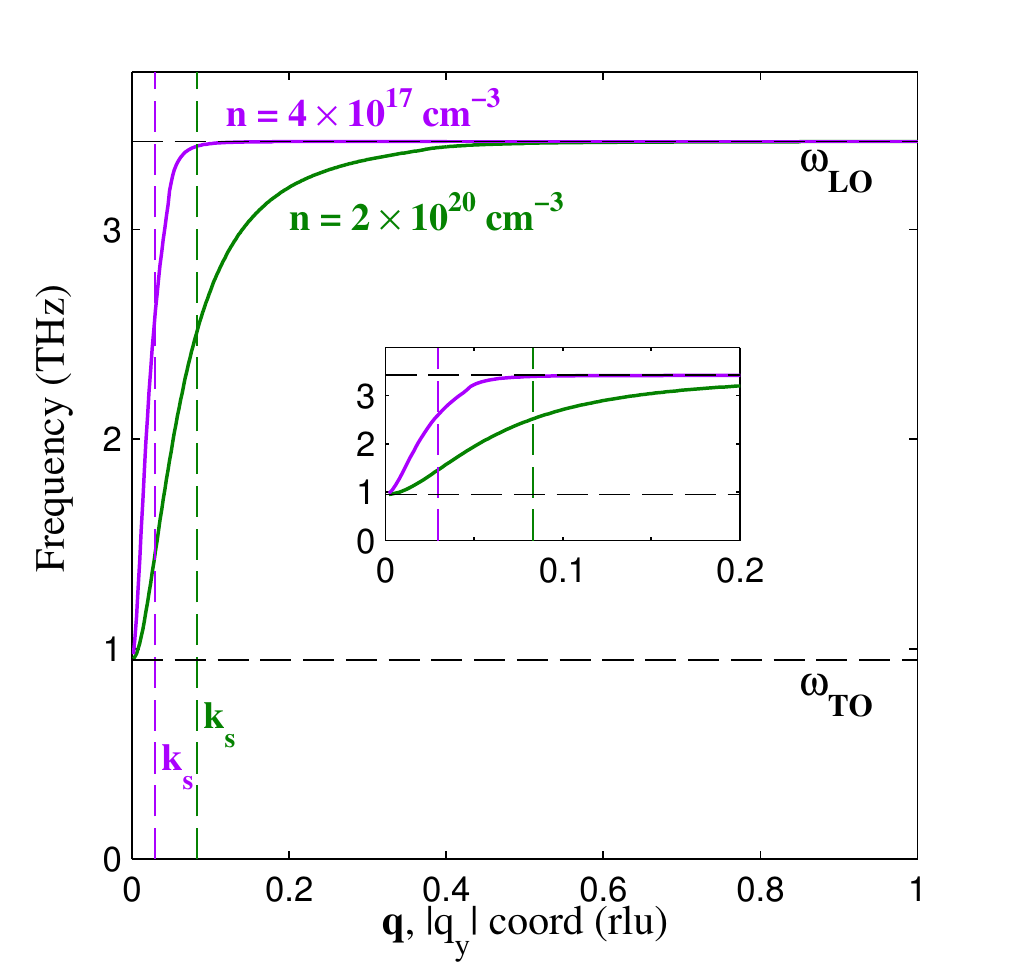}
\caption{\label{f:fig1}Comparison of the calculated dispersion of the screened LO phonon branch along the $\Gamma$ to X direction under carrier densities of $n_c = 2\times 10^{20} \text{ cm}^{-3}$ (green line) and $n_c = 4 \times 10^{17} \text{ cm}^{-3}$ (purple line). The higher carrier density matches the estimated photoexcited carrier density for this experiment, while the lower carrier density represents the initial concentration of the PbTe sample. The dashed black horizontal lines represent the zone center TO and LO frequencies of PbTe. The dashed colored vertical lines are the calculated Thomas-Fermi screening wavevectors. The inset figure plots the same dispersions in a shorter wavevector range in order to emphasize the difference.}
\end{figure}
 
In this model, the calculated dispersion in $\omega_{-}$ at both equilibrium and photo-excited carrier density of $n_c = 4\times 10^{17}\text{cm}^{-3}$ and $2\times 10^{20}\text{cm}^{-3}$ is shown in Figure \ref{f:fig1}, purple and green traces respectively.
In the photoexcited case, $k_s \sim 2.7$ times larger than that for the lower density even though the change in carrier density is more than three orders of magnitude, reflecting the $n_{c}^{1/6}$ dependence  in eq.\ref{e:ks}.  Nonetheless a dramatic shift in the dispersion is observed at low $q$.
Due to the low carrier concentration of our samples, the expected dispersion deviates appreciably from the LO frequency only in a small region near $q=0$.
This is reflective of the relatively weak screening of the LO phonon mode at such low carrier densities and the screened portion of the dispersion would be difficult to resolved in most measurement. On the other hand for the dense photoexcited case the screened region of the dispersion of the LOPC mode extends considerably further from the zone center, reflective of the decreased screening length (increased $k_s$). 
	
	The data in Figure \ref{f:fig2}c and d extend to approximately the $k_s$ calculated above, and thus is expected to disperse rapidly towards the unscreened LO phonon frequency. 
	Thus, we expect that the the LOPC mode $\omega_{-}(\textbf{q})$ should appear prominently in this region, given the strong sensitivity to longitudinal phonon polarization along this wavevector direction. The blue trace overlaid on the FT-IXS spectrum in the figure is obtained by using the reduced wavevector coordinates along $\Gamma$ to W in this experiment in eq. 1. This represents the computed low-$\textbf{q}$ screened LO phonon dispersion. A close agreement with the highly-dispersive spectral feature observed in the FT-IXS experiment is achieved for the estimated photoexcited carrier density. Thus, we attirbute the broad high frequency dispesrsive feature to the photo-induced LOPC mode.{\par}
    We note that in the FT-IXS scheme, this LOPC mode could appear at either its predicted singular frequency or in combination with low energy acoustic modes considering the noted generation of correlated pairs of modes. Examining further, upon photoexcitation of carriers the sudden change of the dielectric function $\varepsilon(\mathbf{q},\omega)$ alters the combination LOPC mode frequencies and eigenvectors\cite{jiang2016natcomm}. Subsequently, the correlation of the modes in the low-frequency continuum of plasma excitations with the LO mode gives rise to a squeezing of its polarization vector. This results in oscillations within the x-ray diffuse scattering at frequencies close to the screened LO phonon frequency. However, coupling of the LO and LA phonon degrees of freedom can also be included, as in the case of pure phonon squeezing, resulting in similar oscillations.{\par}
    Although the sudden excitation of a carrier plasma substantially alters the frequency of the LO modes, a detailed calculation of the resulting plasmon-phonon squeezing shows that this does not give rise to squeezing of the ionic vibrations at very long wavelengths, where $k \ll  k_{F}$ and $\omega_{TO}/\omega_{F} \gg  2k/k_{F}$. However, for somewhat shorter wavelengths, the presence of electron-hole excitations at very low frequency, corresponding to a non-zero imaginary part of the dielectric function for $0 < {\omega}/{\omega_{F}} < \left(2-{k}/{k_{F}}\right){k}/{k_{F}}$,
gives rise to squeezing oscillations at the screened LO phonon frequency, $\omega_{-}(\mathbf{q})$, and corresponding oscillations in the diffuse x-ray scattering, consistent with the experimental observations. Although a renormalization of short-range interatomic force constants in the photoexcited system could in principle lead to phonon squeezing at the combination mode frequencies, $\omega_{LO}$ $\pm$ $\omega_{LA}$, our constrained density functional theory calculations indicate that this squeezing signal is negligible, compared with the squeezing that arises from the coupling of the ionic motion with low frequency electron-hole excitations in the carrier plasma, for wavevectors in the range where the strongly dispersive feature is observed.
{\par}
	The observed photo-induced plasmon-phonon state and the resulting screening of the LO phonon branch are consistent with INS data on PbTe at degenerate carrier concentrations\cite{cowley1965prl,alperin1972pla,cochran1966proc}. The difference is that in the current experiment, we measure the screened LO mode dispersion in much lower equilibrium densities, suddenly excited to  a higher carrier concentration ( the reported INS studies measured samples with relatively high carrier concentrations ($\sim 2\times 10^{19} \text{ cm}^{-3}$ in Cowley and Dolling\cite{cowley1965prl} and $\sim 4\times 10^{18} \text{ cm}^{-3}$ in Alperin $\textit{et al.}$\cite{alperin1972pla}), whereas the PbTe sample examined here starts at a much lower concentration $(4\times 10^{17} \text{ cm}^{-3})$ prior to photoexcitation). In the photo-excited state, the sample naively may be considered in a similar regime as the highly-doped crystals measured with INS, ahtough we stress here that the current measurements do not reflect the spontaneous scattering from single LOPC and other excitation, but a nonthermal and nonstationary state produced by the sudden-excitaiton. 
	this results in the measurement of correlated phonon pairs, for example in the TA overtone and TA$\pm$TO combination modes seen in ref.  \cite{jiang2016natcomm} to span the entire Brillouin zone.
	Here we have shown further that a high photoexcited carrier density in PbTe substantially affects the largely unscreened, LO modes by sudden modification of the screening generation of nonthermal LOPC modes of PbTe.{\par}
    
    This work is supported by the Department of Energy, Office of Science, Basic Energy Sciences, Materials Sciences and Engineering Division, under Contract DE-AC02-76SF00515. I. Savi\'{c} acknowledges support by Science Foundation Ireland and Marie Curie Action COFUND under Starting Investigator Research Grant 11/SIRG/E2113. S. Fahy and \'{E}.D. Murray acknowledge support by Science Foundation Ireland under Grant No. 12/1A/1601. O. Delaire acknowledges support from the U.S. Department of Energy, Office of Science, Basic Energy Sciences, Materials Sciences and Engineering Division, through the Office of Science Early Career Research Program. 
    Sample synthesis (AFM, BCS) was supported by  the U.S. Department of Energy, Office of Science, Basic Energy Sciences, Materials Sciences and Engineering Division
    J.C. acknowledges financial support from the Volkswagen Foundation. Portions of this research were carried out at the Linac Coherent Light Source (LCLS) at the SLAC National Accelerator Laboratory. LCLS is an Office of Science User Facility operated for the U.S. Department of Energy Office of Science by Stanford University. Preliminary experiments were performed at SACLA with the approval of the Japan Synchrotron Radiation Research Institute (JASRI) (Proposal No. 2013A8038) and at the Stanford Synchrotron Radiation Lightsource, SLAC National Accel-erator Laboratory, which like the LCLS is supported by the U.S. Department of Energy, Office of Science, Office of Basic Energy Sciences under Contract No. DE AC02-76SF00515.{\par}

\bibliography{plasmon}
\end{document}